\documentclass[amsmath,amssymb,superscriptaddress,nobalancelastpage,prl,twocolumn]{revtex4-1}

\usepackage{graphicx}
\usepackage{varioref}
\usepackage{xr-hyper}
\usepackage{xcolor}
\usepackage{hyperref}
\hypersetup{colorlinks,linkcolor=blue,urlcolor=blue,citecolor=blue}
\usepackage{ulem}
\usepackage{braket}
\newcommand{\red}[1]{{\textcolor{black}{#1}}}

\newcommand{\LSCO}{La$_{2-x}$Sr$_x$CuO$_4$}

\newcommand{\Hgs}{HgBa$_2$CuO$_{4+\delta}$}

\newcommand{\Tc}{$T_\mathrm{c}$}
\newcommand{\Hgt}{HgBa$_2$Ca$_2$Cu$_3$O$_{8+\delta}$}
\newcommand{\HgRet}{(Hg,Re)Ba$_2$Ca$_2$Cu$_3$O$_{8+\delta}$}
\newcommand{\Bis}{(Bi,Pb)$_2$(Sr,La)$_2$CuO$_{6+\delta}$}
\newcommand{\Bib}{Bi$_2$Sr$_2$CaCu$_2$O$_{8+\delta}$}
\newcommand{\Bit}{Bi$_2$Sr$_2$Ca$_2$Cu$_3$O$_{10+\delta}$}
\newcommand{\EF}{$E_\mathrm{F}$}

\newcommand{\YBCO}{YBa$_2$Cu$_3$O$_{7- \delta}$}

\newcommand{\dz}{$d_{3z^2-r^2}$}
\newcommand{\dx}{$d_{x^2-y^2}$}

\begin{document}
\author{M.~Horio}
	\email{mhorio@issp.u-tokyo.ac.jp}
 	\affiliation{Institute for Solid State Physics, The University of Tokyo, Kashiwa, Chiba 277-8581, Japan}

\author{M.~Miyamoto}
	\affiliation{Institute for Solid State Physics, The University of Tokyo, Kashiwa, Chiba 277-8581, Japan}
	
\author{Y.~Mino}
	\affiliation{Department of Physics, Tokyo University of Science, Shinjuku, Tokyo 162-8601, Japan}
	\affiliation{National Institute of Advanced Industrial Science and Technology (AIST), Tsukuba, Ibaraki 305-8568, Japan}

\author{S.~Ishida}
	\affiliation{National Institute of Advanced Industrial Science and Technology (AIST), Tsukuba, Ibaraki 305-8568, Japan}

\author{B.~Thiagarajan}
	\affiliation{Max IV Laboratory, Lund University, Box 118, 22100 Lund, Sweden}

\author{C.~M.~Polley}
	\affiliation{Max IV Laboratory, Lund University, Box 118, 22100 Lund, Sweden}

\author{C.~H.~Lee}
\affiliation{National Institute of Advanced Industrial Science and Technology (AIST), Tsukuba, Ibaraki 305-8568, Japan}

\author{T.~Nishio}
\affiliation{Department of Physics, Tokyo University of Science, Shinjuku, Tokyo 162-8601, Japan}

\author{H.~Eisaki}
	\affiliation{National Institute of Advanced Industrial Science and Technology (AIST), Tsukuba, Ibaraki 305-8568, Japan}
	
\author{I.~Matsuda}
	\affiliation{Institute for Solid State Physics, The University of Tokyo, Kashiwa, Chiba 277-8581, Japan}

%\title{Enhanced outer-plane superconducting gap in the trilayer cuprate \HgRet}

\title{Enhanced superconducting gap \red{in the outer CuO$_2$ plane of} the trilayer cuprate \HgRet}

\begin{abstract}
We report the first observation of a momentum-resolved superconducting gap in the Hg-based trilayer cuprate, which holds the highest record of superconducting transition temperature (\Tc) at ambient pressure. By  angle-resolved photoemission spectroscopy utilizing a micro-focused beam, clear quasiparticle dispersions originating from the inner and outer CuO$_2$ planes \red{(IP and OP, respectively)} were separately identified. The magnitude of the superconducting gap for the IP was comparable to that of the Bi-based trilayer cuprate with a lower \Tc. In contrast, the superconducting gap for the OP \red{was significantly larger than that of the Bi-based one}. While strong pairing in the IP has been highlighted as the key element of trilayer cuprates, the present results suggest that \red{the enhanced pairing energy in the OP is essential for the highest \Tc\ at ambient pressure realized} in the Hg-based trilayer cuprate.
 %enhancing the paring strength in the OP is essential for boosting \Tc\ to the highest value in the Hg-based trilayer cuprates.

\end{abstract}

\maketitle

\begin{figure}[t]
	\begin{center}
		\includegraphics[width=0.48\textwidth]{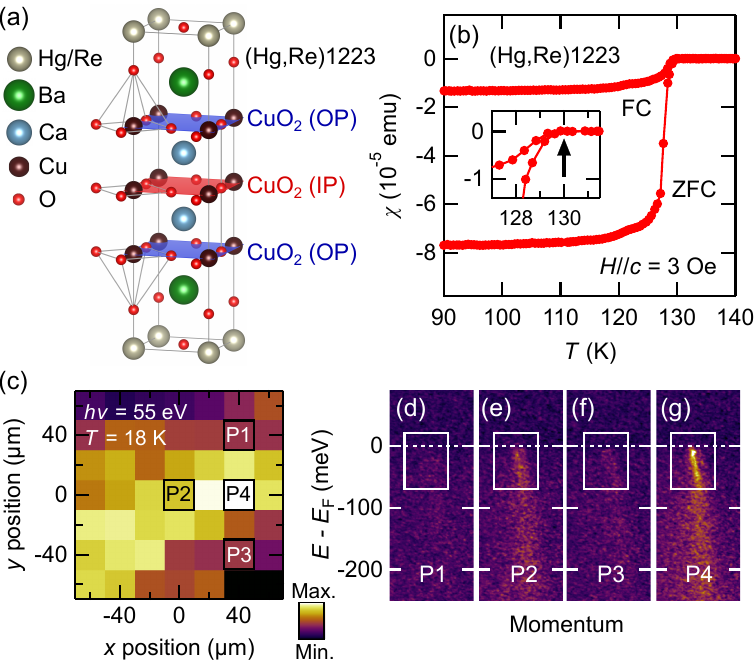}
	\end{center}
	\caption{\textbf{Preparation for the ARPES study of (Hg,Re)1223.} \red{(a) Crystal structure of (Hg,Re)1223~\cite{VESTA}.} \red{(b)} Magnetic susceptibility of a (Hg,Re)1223 single crystal signifying the \Tc\ of 130~K. FC and ZFC denote field and zero-field cooling, respectively. \red{(c)} Real-space map of photoemission intensity from nodal quasiparticles. The map covers the region of 120~$\mu$m square within the $\sim 500 \mu$m square sized sample. \red{(d)-(g)} Energy-momentum images acquired at sample positions P1--P4, respectively, indicated in \red{(c)}. The photoemission intensity from nodal quasiparticles integrated within the white rectangles is displayed in \red{(c)}.} 
	\label{pos}
\end{figure}

%Introduction
%\section{Introduction}
The material dependence of superconducting transition temperature (\Tc) provides useful insights into the mechanism for the high \Tc\ of cuprate superconductors. The number of CuO$_2$ layers in a unit cell, $n$, has been known as one of the main control parameters. \Tc\ is enhanced with increasing $n$ and maximized at $n=3$, followed by a moderate reduction at larger $n$~\cite{ParkinPRL1988,TarasconPRB1988,IyoJPSJ2007}. The key feature of the $n=3$, trilayer cuprates is the existence of two inequivalent CuO$_2$ planes, the inner plane (IP) and outer plane (OP) \red{[Fig.~\ref{pos}(a)]}. Among trilayer cuprates, \Bit\ (Bi2223) with the optimal \Tc\ of 110~K~\cite{TarasconPRB1988} has been intensively studied by angle-resolved photoemission spectroscopy (ARPES) owing to the ease of cleaving, and quasiparticle dispersions originating from the IP and OP were separately observed. The IP (OP) was found to be underdoped (overdoped) with a larger (smaller) superconducting gap~\cite{IdetaPRL2010,IdetaPRB2012,KunisadaPRL2017,IdetaPRL2021,LuoNatPhys2023}. As such, the large pairing \red{energy} in the IP has been suggested essential for the high \Tc~\cite{KunisadaPRL2017,LuoNatPhys2023}, while the overdoped OP with large phase stiffness should also be involved through interlayer interactions such as proximity effects~\cite{KivelsonPhysicaB2002,BergPRB2008,OkamotoPRL2008} and pair hopping~\cite{NishiguchiPRB2013} or scattering~\cite{NishiguchiPRB2018}.

Even within the trilayer cuprate family, substantial \Tc\ variations are found. The highest value of 134~K is realized in \Hgt\ (Hg1223)~\cite{SchillingNature1993}, which is also known as the highest record of \Tc\ at ambient pressure among all the existing superconductors. %thus identifying the origin of the enhanced \Tc\ has been a central issue. %Bi2223 and Hg1223 share the same CuO$_5$(OP)-Ca-CuO$_4$(IP)-Ca-CuO$_5$(OP) unit.
Key differences between Bi2223 and Hg1223 have been explored mostly from the structural perspective. Whereas in the trilayer cuprates the IP is free from buckling and shielded from out-of-plane disorder, the OP is generally buckled and directly exposed to disorder potential. Yet, the buckling angle evaluated by diffraction measurements is one order of magnitude smaller in Hg1223 than in Bi2223~\cite{WagnerPRB1995}. In addition, the known source of out-of-plane disorder is located farther from the OP in Hg1223 than in Bi2223~\cite{EisakiPRB2004}. \red{Such} high degrees of flatness and cleanliness of the OP in Hg1223 were also captured by nuclear-magnetic-resonance (NMR) measurements from the sharpness of the resonance peak~\cite{MagishiJPSJ1995,IwaiJPSCP2014}. Despite thorough structural charaterization, however, little is yet resolved on the low-energy electronic structure of Hg1223 that has direct relevance to superconductivity. %If a direct comparison to the known electronic structure of Bi2223 had been possible, the key aspect that is responsible for the record-high \Tc\ would have been unambiguously identified.
Investigations \red{into its electronic structure} by ARPES have been hampered by the limited availability of single crystals. \red{Moreover, the absence of natural cleavage planes in Hg1223 adds another layer of difficulty in performing ARPES measurements. } Unlike \red{the case of} Bi2223, \red{cleaving a} Hg1223 \red{crystal} should yield a disordered surface \red{whose randomness is problematically mirrored in ARPES spectra}.

To make a breakthrough, Mino \textit{et al.}~\cite{MinoJPSJ2024} have recently established methodology to grow single crystals of the Hg-based trilayer cuprate with high reproducibility through partial Re substitutions, and single-crystalline \HgRet\ [(Hg,Re)1223] samples with \Tc's exceeding 130~K have been reproducibly synthesized. Although inhomogeneity created by cleaving would be inevitable, one could purify ARPES signals if the beam is highly focused and thus the probing area is sufficiently small. 

In this Letter, we uncover momentum-resolved electronic structure of single-crystalline (Hg,Re)1223 with $T_\mathrm{c}=130$~K by \red{utilizing} micro-spot ARPES. Although the appearance of ARPES spectra drastically varied over space, Fermi surfaces and superconducting gaps were captured in the CuO$_2$-layer specific manner by exploiting the small beam spot size of 10~$\mu$m$\times$10~$\mu$m. The magnitude of the superconducting gap was found to be comparable to that of optimally doped Bi2223 for the IP, but substantially \red{larger than that of Bi2223} for the OP. The results establish the \red{large pairing energy} in the OP as a key ingredient for the highest \Tc\ at ambient pressure in the Hg-based trilayer cuprates.

%\section{Methods}
\red{Bulk} single crystals of (Hg,Re)1223 \red{with the in-plane dimensions of $\sim 500$ $\mu\mathrm{m} \times 500$ $\mu\mathrm{m}$ and the thickness of $\sim 40$ $\mu\mathrm{m}$} were synthesized by the self-flux method~\cite{MinoJPSJ2024}. \red{The actual Hg:Re ratio was close to the nominal composition (0.8:0.2).} The \Tc\ determined from the onset of the Meissner signal was 130~K [Fig.~\ref{pos}\red{(b)}], suggesting a nearly optimal hole-doping level. ARPES measurements were performed at the Bloch beamline of MAX IV~\cite{Bloch} with a beam focused to 10~$\mu$m$\times$10~$\mu$m. Single crystals were cleaved \textit{in-situ} and measured at $T=18$~K under a vacuum better than $1 \times 10^{-10}$~Torr. The incident photon energy and total energy resolution were set at $h\nu=55$~eV and $\Delta E = 12$~meV, respectively.

%\section{Results and Discussion}
%position dependence
Before performing detailed investigations into the electronic structure, the optimal sample position for ARPES measurements has been explored. Figure~\ref{pos}\red{(c)} shows a two-dimensional real-space map of photoemission intensity from the nodal quasiparticles integrated within the white boxes in Figs.~\ref{pos}\red{(d)-(g)}. The quasiparticle peak intensity varies pixel by pixel, where a pixel covers 20~$\mu$m $\times$ 20~$\mu$m. While a clear nodal quasiparticle dispersion is observed at position P4 [Fig.~\ref{pos}\red{(g)}], the spectra lose clarity upon moving away by 40~$\mu$m [Figs.~\ref{pos}\red{(d)-(g)}]. Yet, with the spot size of 10~$\mu$m$\times$10~$\mu$m, it was possible to pinpoint the optimal area and obtain ARPES spectra of the highest quality.

\begin{figure*}[ht]
	\begin{center}
		\includegraphics[width=\textwidth]{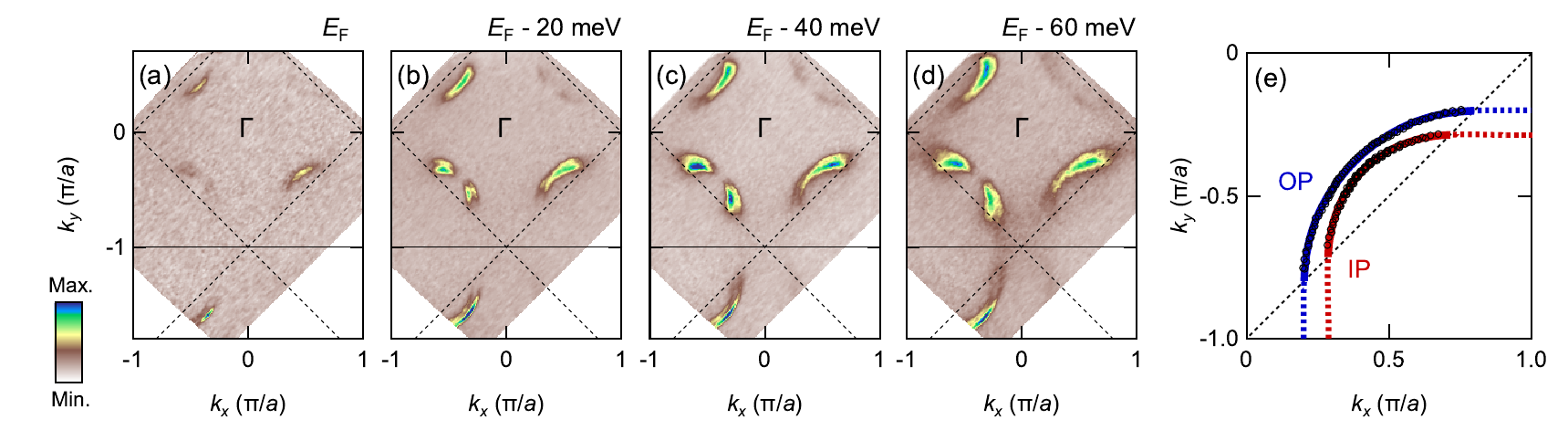}
	\end{center}
	\caption{\textbf{Fermi surface of (Hg,Re)1223.} (a)-(d) Constant-energy surfaces constructed by integrating photoemission intensity within $\pm 10$~meV around the indicated energies. (e) Fermi momenta $k_\mathrm{F}$'s (black circles) \red{gathered to the quarter of the BZ according to the crystalline symmetry and } fitted by the tight-binding model. The fitted curve for the IP and OP are indicated in red and blue, respectively.} 
	\label{FS}
\end{figure*}

%Fermi surface
Fixing the sample position at P4 in Fig.~\ref{pos}, constant energy surfaces were mapped in $k_x$-$k_y$ space [Figs.~\ref{FS}(a)-(d)]. Only the nodal part is appreciable at \EF, and the arc extends toward the antinodal region at deeper energies, reflecting the opening of a $d$-wave-like gap. Still, the spectral intensity is strongly suppressed once going beyond the $\sqrt{2} \times \sqrt{2}$ antiferromagnetic Brillouin-zone (BZ) boundary due probably to pseudogap opening. We determined Fermi momentum $k_\mathrm{F}$ from the gap minimum at each cut and plotted the position % as a black circle
in Fig.~\ref{FS}(e). Two branches, representing the IP and OP, were found as is the case for optimally doped Bi2223~\cite{IdetaPRL2010,IdetaPRB2012,KunisadaPRL2017,IdetaPRL2021}. \red{Referring to the previous NMR studies~\cite{MagishiJPSJ1995,IwaiJPSCP2014} that reported smaller hole concentrations in the IP than in the OP, the red (blue) branch in Fig.~\ref{FS}(e) with the smaller (larger) radius was assigned to the IP (OP).} We fitted these Fermi surfaces to the following tight-binding model: $\epsilon-\mu = \epsilon_0 - 2t[\cos(k_xa)+\cos(k_ya)] - 4t'\cos(k_xa)\cos(k_ya) - 2t''[\cos(2k_xa)+\cos(2k_ya)]$, where $t$, $t'$, and $t''$ represent nearest, second-nearest, and third-nearest neighbor hopping parameters, respectively, and $\epsilon_0$ is the center of the band relative to the chemical potential $\mu$. Fixing $t''/t'$ at $-0.5$ as frequently assumed~\cite{YoshidaPRB2006,HashimotoPRB2008,IdetaPRL2010}, we obtained $t'/t = -0.36$ and $\epsilon_0 /t = 0.40$ for the IP and $t'/t = -0.28$ and $\epsilon_0 /t = 0.91$ for the OP. Area of the fitted Fermi surface yields the hole doping level $p_\mathrm{FS}$ of -0.06 and 0.12 for the IP and OP, respectively, where the negative sign indicates electron doping. However, as the pseudogap in the electron-doped case should open around the node instead of the antinode~\cite{ArmitageRMP2010,HorioPNAS2025}, electron doping in the IP is unlikely. Furthermore, the difference in the carrier concentrations between the IP and OP, $\Delta p$, amounts to 0.18, which is considerably larger than $\Delta p = 0.05$ estimated in the previous NMR studies~\cite{MagishiJPSJ1995,IwaiJPSCP2014}.
%Considering one IP and two OPs in a unit cell, the averaged hole concentration per CuO$_2$ plane is 0.06. This number also deviates from an expectation for optimally doped cuprates which usually have $p \sim 0.16$~\cite{KeimerNat2015}. These inconsistencies imply the breakdown of the Luttinger's theorem under the assumption of the large Fermi surface centered at ($\pi$, $\pi$).
Inconsistencies associated with the underestimate of hole concentrations were also reported for underdoped Bi$_2$(Sr,La)$_2$CuO$_{6+\delta}$~\cite{HashimotoPRB2008} and (Ca,Na)$_2$CuO$_2$Cl$_2$~\cite{ShenScience2005} and reconciled by exploiting the doped resonant valence-bond spin-liquid concept~\cite{YangPRL2011,MengPRB2011}, where a small Fermi pocket formed within the original Fermi surface and the antiferromagnetic BZ boundary is proposed to be a more appropriate description in the pseudogap state~\cite{YangPRB2006}. Recent results of Hall-effect measurements indeed revealed a broad transition from a small pocket to a large Fermi surface across the psuedogap critical point~\cite{BadouxNature2016,CollignonPRB2017,PutzkeNatPhys2021}. 

%small Fermi surface
Based on the small Fermi pocket picture, we re-evaluated carrier concentrations. As determining the exact shape of the Fermi surface is challenging, an estimate was made by taking average of the two extreme cases that can be quantitatively defined~\cite{Supplemental}. This approximation was recently applied to \red{a} La-based cuprate \red{with the so-called T*-type crystal structure} and successfully yielded $p_\mathrm{FS}$ consistent with the nominal value~\cite{HorioPRB2023}. The resulting $p_\mathrm{FS}$ is 0.09 and 0.17 for the IP and OP, respectively. 
%$\Delta p_\mathrm{FS}$ is now reduced to 0.08, which is closer to the NMR estimate.
%The averaged $p_\mathrm{FS}$ of 0.15 per CuO$_2$ plane is also reasonable for a nearly optimally doped cuprate.
The small pocket picture is thus necessary for the IP to be hole doped. On the other hand, its applicability to the OP with a higher hole doping level is elusive. $\Delta p_\mathrm{FS}$ does not provide a definitive clue to solving this issue as it is equally close to the NMR estimate of 0.05~\cite{MagishiJPSJ1995,IwaiJPSCP2014} when the small and large Fermi surface scenarios are applied to the OP ($\Delta p_\mathrm{FS}=0.08$ and 0.03, respectively). Nevertheless, it is safely concluded -- irrespective of the assumptions -- that the OP of (Hg,Re)1223 [$p_\mathrm{FS}\mathrm{(OP)} = 0.12$--0.17] is less doped than that of optimally doped Bi2223 [$p_\mathrm{FS}\mathrm{(OP)} = 0.23$]~\cite{IdetaPRL2010}. These observations, in turn, suggest that surface carrier reconstruction observed at the polar surface of \YBCO~\cite{NakayamaPRB2007,ZabolotnyyPRB2007,IwasawaPRB2018} is unlikely for (Hg,Re)1223. Since CuO$_2$ planes are negatively charged, any reconstruction to compensate surface polarity should lead to hole overdoping into the CuO$_2$ planes, which is not the present case. The cleaved surface of the single-layer counterpart \Hgs\ (Hg1201) is also known to avoid such reconstruction possibly by making an equal amount of positively and negatively charged terminations, yielding an overall neutral surface~\cite{VishikPRB2014,VishikPRB2020}. A similar situation would be realized at the surface of (Hg,Re)1223.

\begin{figure}[t]
	\begin{center}
		\includegraphics[width=0.49\textwidth]{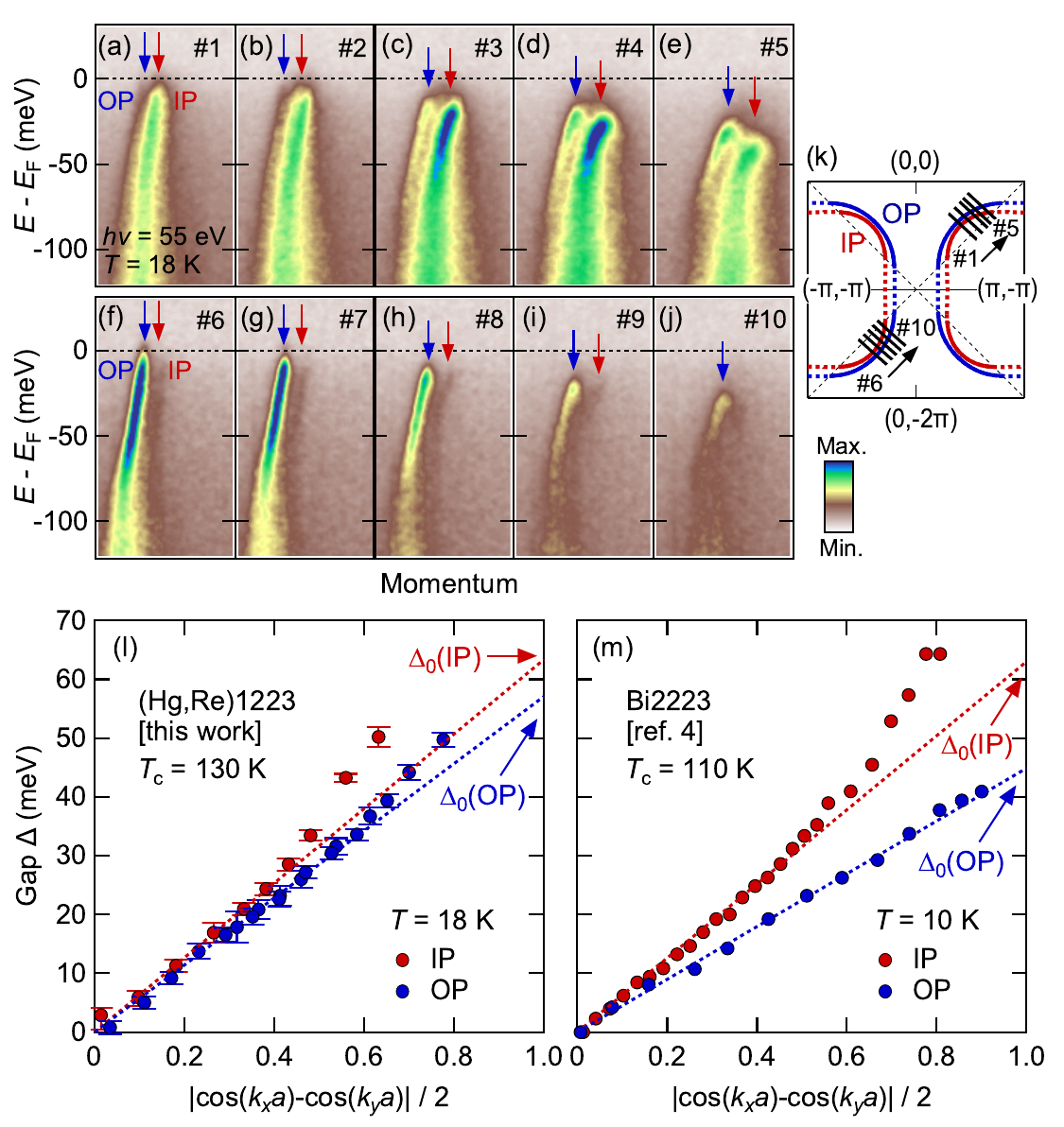}
	\end{center}
	\caption{\textbf{Momentum-dependent energy gap opening.} (a)-(j) Energy-momentum images along cuts \#1-\#\red{10}, respectively, indicated in panel \red{(k)}. The IP and OP bands are marked by the red and blue arrows, respectively. \red{(k)} The positions of the momentum cuts. \red{(l)} The magnitude of the gap $\Delta$ plotted against the $d$-wave order parameter $|$cos($k_x a$)-cos($k_y a$)$|/2$. The dashed lines indicate the extrapolation of the simple $d$-wave from to obtain $\Delta_0$. \red{(m) The same as (l) but for optimally doped Bi2223 taken from ref.~\onlinecite{IdetaPRL2010}.}} 
	\label{gap}
\end{figure}

%gap
With the Fermi surface characterized, the momentum dependence of the spectral gap shall be examined. Around the node in the first BZ [Fig.~\ref{gap}(a)], photoemission signals from the IP are dominant. Upon moving toward the antinode, the signals from the OP gain intensity, and clear spectral gaps open both for the IP and OP [Figs.~\ref{gap}(b)-\red{(e)}]. On the other hand, in the second BZ [Figs.~\ref{gap}\red{(f)-(j)}], the spectral intensity from the OP is persistently enhanced due to the matrix-element effect acting differently. We thus estimated gap magnitude $\Delta$ for the IP at the first BZ, and combined information from the first and second BZs for the OP.
%While the gap magnitude $\Delta$ can be determined directly from the energy position of the coherent peak, the estimate becomes unreliable when the size is comparable or smaller than the energy resolution. In such cases, specifically in the nodal region where the $d$-wave order parameter $|$cos($k_x a$)-cos($k_y a$)$|/2 < 0.2$, we estimated $\Delta$ from the leading-edge shift $\Delta_\mathrm{LEM}$, employing the empirical relationship of $\Delta \sim 2.2\Delta_\mathrm{LEM}$~\cite{YoshidaPRL2009}.
Evaluated $\Delta$ values~\cite{Supplemental} are plotted in Fig.~\ref{gap}\red{(l)} both for the IP and OP as a function of the $d$-wave order parameter. While $\Delta$ increases proportionally near the node, a deviation is found in the antinodal region. This is typical behavior of cuprates in the underdoped to optimally doped regime~\cite{TanakaScience2006,KondoPRL2007,KondoNature2009,YoshidaPRL2009,HashimotoPRB2009,VishikPNAS2012}. \red{Since the antinodal gap deviating from the simple $d$-wave form remains open above \Tc\ in those cuprates, it has been regarded as a pseudogap~\cite{KondoPRL2007,KondoNature2009,YoshidaPRL2009,VishikPNAS2012}. To separate the superconducting gap from the pseudogap, in the previous studies, the near-nodal simple $d$-wave part was extrapolated to $|$cos($k_x a$)-cos($k_y a$)$|/2 = 1$ and the gap $\Delta_0$ was defined. $\Delta_0$ is known to be virtually independent of doping levels around the optimal doping~\cite{YoshidaPRL2009,VishikPNAS2012}. Instead, $\Delta_0$ differs for different cuprate systems and scales with the maximal \Tc\ of the material at the optimal doping~\cite{TanakaScience2006,KondoNature2009,YoshidaPRL2009,VishikPNAS2012,VishikPRB2014}, demonstrating its nature as a superconducting gap.}  %Under the coexistence of the pseudogap, frequently applied  analysis to extract superconducting properties is to define $\Delta_0$ by extrapolating the simple $d$-wave part to $|$cos($k_x a$)-cos($k_y a$)$|/2 = 1$. $\Delta_0$ has been known to be material dependent reflecting the strength of pairing, but virtually independent of doping levels around the optimal doping~\cite{YoshidaPRL2009,VishikPNAS2012}.
In the present case, we obtained $\Delta_0$(IP) = $63 \red{ \pm 3}$~meV and $\Delta_0$(OP) = $57 \red{ \pm 1}$~meV for (Hg,Re)1223 \red{ [Fig.~\ref{gap}(l)], where the error bar represents 90\% confidence interval of the linear extrapolation}.

It is constructive to track changes of $\Delta_0$ from the bilayer compound \Bib\ (Bi2212) through Bi2223 to (Re,Hg)1223 around the optimal doping. Bi2212 forms two distinct Fermi surfaces arising from bonding and antibonding hybridization of the two equivalent CuO$_2$ planes, and their $\Delta_0$ values nearly degenerate~\cite{AiCPL2019} at $\sim 40$~meV~\cite{TanakaScience2006,LeeNature2007,VishikPNAS2012}. As for Bi2223 with another CuO$_2$ plane IP inserted, $\Delta_0$(OP) $\sim 43$~meV remains comparable to the $\Delta_0$ of Bi2212, but $\Delta_0$(IP) is drastically enhanced to $\sim 60$~meV~\cite{IdetaPRL2010}. The underdoped IP should suffer from strong phase fluctuations due to low superfluid stiffness, and the \Tc\ would be suppressed despite the large pairing energy scale if the CuO$_2$ plane were isolated~\cite{UemuraPRL1989}. In Bi2223, however, the adjacent overdoped OP may serve as a complementary source of phase stiffness through %coupling to the IP
interlayer coupling to make full use of the pairing \red{energy} at the IP~\cite{KivelsonPhysicaB2002,BergPRB2008}. Such a composite picture to realize a high \Tc\ was supported by the recent ARPES study on overdoped Bi2223 by Luo \textit{et al.}~\cite{LuoNatPhys2023}. It has been known that the \Tc\ of Bi2223 in the overdoped region is maintained at the optimal value independent of the doping level~\cite{FujiiPRB2002}. For the overdoped Bi2223 sample with the optimal \Tc, Luo \textit{et al.}~\cite{LuoNatPhys2023} found a decrease of $\Delta_0$(OP) from the optimally doped value, while $\Delta_0$(IP) remained unchanged. Thus, the dominant role of $\Delta_0$(IP) has been suggested in determining the \Tc\ of trilayer cuprates~\cite{KunisadaPRL2017,LuoNatPhys2023}. Turning attention to the variation within the trilayer cuprate family, however, $\Delta_0$(IP) of (Hg,Re)1223 (63\red{$\pm3$}~meV)  remains comparable to that of Bi2223~\cite{IdetaPRL2010} \red{[see the comparative plots in Figs.~\ref{gap}(l) and (m)]} despite the increase of \Tc\ by $\sim 20$~\% at the optimal doping. In contrast, $\Delta_0$(OP) of (Hg,Re)1223 (57\red{$\pm1$}~meV) is substantially enhanced from that of optimally doped Bi2223~\cite{IdetaPRL2010} \red{as one can see from Figs.~\ref{gap}(l) and (m)}. The difference in the optimal \Tc\ between Bi2223 and (Hg,Re)1223 is thus rooted apparently in $\Delta_0$(OP) rather than in $\Delta_0$(IP).

\begin{figure}[t]
	\begin{center}
		\includegraphics[width=0.4\textwidth]{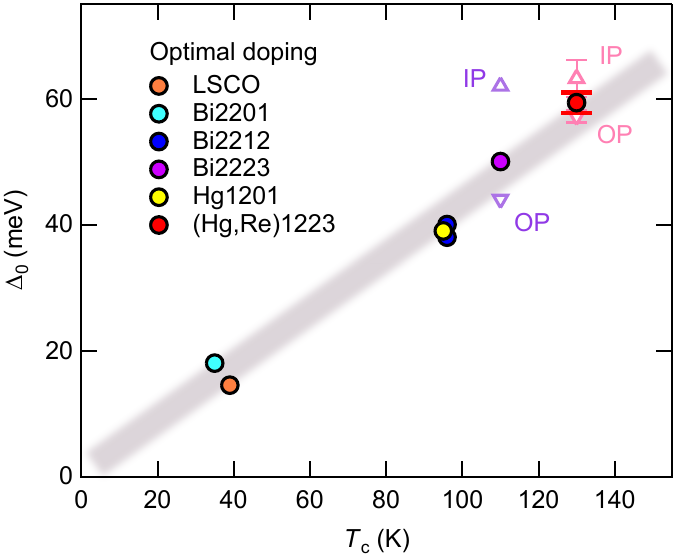}
	\end{center}
	\caption{\textbf{Relation between $\Delta_0$ and \Tc\ at the optimal doping.} The data for \LSCO\ (LSCO)~\cite{YoshidaPRL2009}, \Bis\ (Bi2201)~\cite{KondoNature2009}, Bi2212~\cite{TanakaScience2006,VishikPNAS2012}, Bi2223~\cite{IdetaPRL2010}, and Hg1201~\cite{VishikPRB2014} have been taken from previous ARPES studies. The individual $\Delta_0$ values for the IP and OP are plotted for the trilayer cuprates Bi2223 and (Hg,Re)1223. The grey line is for a guide to the eyes.} 
		\label{Delta0}
	\end{figure}

As shown in Fig.~\ref{Delta0}, $\Delta_0$ of single-layer and bilayer optimally doped cuprates~\cite{YoshidaPRL2009,KondoNature2009,VishikPNAS2012,VishikPRB2014} displays an intimate link to the optimal \Tc. As for the trilayer cuprates, the largest $\Delta_0$ of the system, $\Delta_0$(IP), deviates from the trend set by the single-layer and bilayer cuprates; the increase from Bi2212 or Hg1201 to Bi2223 is too steep, while the change from Bi2223 to (Re,Hg)1223 is too subtle. Instead, by taking average over the CuO$_2$ planes~\cite{IdetaPRL2010}, a better scaling is found between the optimal \Tc\ and $\Delta_0$ throughout these cuprate families (Fig.~\ref{Delta0}). The result implies that, as far as the optimally doped samples are concerned, the material dependence of \Tc\ in cuprates is governed primarily by the CuO$_2$-plane averaged pairing \red{energy}. The largest value is achieved in the Hg-based trilayer cuprates through   the amplification of pairing at the OP.

Besides the smaller influence of out-of-plane disorder~\cite{EisakiPRB2004}, advantages in the OP of the trilayer Hg-based cuprates can be found from the viewpoint of orbital energy levels and atomic site potential. Due to larger apical oxygen distance from the OP in Hg1223 than in Bi2223~\cite{WagnerPRB1995}, crystal-field splitting between the Cu \dx\ and \dz\ orbitals at the OP should be larger in Hg1223. On the other hand, the Madelung potential difference between the apical and planer (OP) oxygen sites $\Delta V_\mathrm{A}$ was also reported to be larger in Hg1223 than in Bi2223~\cite{JohnstonPRB2010}. $\Delta V_\mathrm{A}$ is correlated with the energy level difference $\Delta E_p$ between the planer O $p_x, p_y$ and apical O $p_z$ orbitals and in turn with the level offset between the \dx\ and \dz\ Wannier orbitals~\cite{SakakibaraPRB2012,SakakibaraPRB2014}. Combined, hybridization between the \dx\ and \dz\ orbitals at the OP, which suppresses $d$-wave superconductivity~\cite{SakakibaraPRL2010,SakakibaraPRB2012,SakakibaraPRB2014}, is expected to be smaller in Hg1223. %, which is compatible with the present $-t'/t$(OP) = -0.28 larger than that of Bi2223 [$-t'/t$(OP) = -0.26]~\cite{IdetaPRL2010}.
Larger $\Delta V_\mathrm{A}$ also implies stronger local electric field along the $c$-axis at the OP planer oxygen sites, which promotes electron-photon coupling for the $c$-axis oxygen-buckling modes~\cite{JohnstonPRB2010}. It was reported based on theoretical analysis that among various phonon modes, coupling to the $B_{1g}$ oxygen-buckling mode could particularly enhance $d$-wave superconductivity~\cite{IdetaPRL2021}. All these elements would collaboratively contribute to reinforcing pairing at the OP in the Hg-based trilayer cuprates. Note that the decrease in the doping level of the OP from Bi2223 to (Hg,Re)1223 might also make an influence on $\Delta_0$ even though $\Delta_0$ should be virtually fixed in a wide doping range~\cite{YoshidaPRL2009,VishikPNAS2012}. A rigorous test can be made on this aspect once ARPES studies of Bi2223 are extended toward the underdoped side.

On the other hand, in the close vicinity of the IP, the pairing \red{energy} at the OP could be enhanced also through the proximity effects~\cite{KivelsonPhysicaB2002,BergPRB2008,OkamotoPRL2008} and/or pair hopping~\cite{NishiguchiPRB2013}/scattering~\cite{NishiguchiPRB2018}. Previous ARPES studies on Bi2223~\cite{KunisadaPRL2017,IdetaPRL2021,LuoNatPhys2023} revealed interlayer hybridization of Bogoliubov quasiparticles arising from interlayer single-particle hopping and/or Cooper pairing. It is intriguing how the interlayer interactions are modulated in the Hg-based trilayer cuprates under the unique environment of the OP. Such hybridization is, however, currently unresolved in (Hg,Re)1223 due likely to the choice of modest energy resolution to capture overall electronic structure efficiently. Now that the feasibility of ARPES studies is demonstrated, thorough investigations into finer spectral features lie ahead to quantify various couplings including the interlayer~\cite{KunisadaPRL2017,IdetaPRL2021,LuoNatPhys2023} as well as the intralayer ones (electron-phonon)~\cite{IdetaJP2013}. Such efforts will pave the way for narrowing down the essential ingredients for the high \Tc\ of the Hg-based trilayer cuprates, and eventually for exploring a path to unlocking the current limitation of \Tc\ at ambient pressure.

%Conclusion
In conclusion, we have performed micro-spot ARPES measurements on (Hg,Re)1223 with $T_\mathrm{c} = 130$~K and observed the quasiparticle dispersions, Fermi surfaces, and superconducting gaps separately for the IP and OP. Hole concentrations were evaluated from the Fermi surfaces, and the OP was found to be less doped than in optimally doped Bi2223.
%The doping level per CuO$_2$ plane, deduced from Fermi surface analysis employing the small pocket picture, was close to the value from other optimally doped cuprates but with an imbalance between the IP and OP.
While the superconducting gap at the IP, $\Delta_0$(IP), was comparable with that of optimally doped Bi2223 with a lower \Tc, \red{$\Delta_0$(OP) was significantly larger than that of Bi2223}. The characteristics of the OP in (Hg,Re)1223 were discussed in terms of the orbital energy levels and atomic site potential. Amplifying the pairing \red{energy} in the OP, and hence on average over the CuO$_2$ planes, is suggested as a key to optimizing \Tc\ in the trilayer cuprates and thus to the highest \Tc\ at ambient pressure \red{realized} in the Hg-based one.

\vspace{5mm}
\begin{acknowledgments}
Fruitful discussion with M.~Kitatani\red{, H.~Suzuki, S.~Smit, M.~Bluschke, and A.~Damascelli} is greatfully acknowledged. This work was supported by JSPS KAKENHI Grant Numbers~JP19H05823 and JP24KK0061. We acknowledge MAX IV Laboratory for time on Beamline Bloch under Proposals 20230779 and 20240484. Research conducted at MAX IV, a Swedish national user facility, is supported by the Swedish Research council under contract 2018-07152, the Swedish Governmental Agency for Innovation Systems under contract 2018-04969, and Formas under contract 2019-02496. 
\end{acknowledgments}

%\bibliography{cuprate}


\end{document}

\author{M.~Horio}
	\email{mhorio@issp.u-tokyo.ac.jp}
 	\affiliation{Institute for Solid State Physics, The University of Tokyo, Kashiwa, Chiba 277-8581, Japan}

\author{M.~Miyamoto}
	\affiliation{Institute for Solid State Physics, The University of Tokyo, Kashiwa, Chiba 277-8581, Japan}
	
\author{Y.~Mino}
	\affiliation{Department of Physics, Tokyo University of Science, Shinjuku, Tokyo 162-8601, Japan}
	\affiliation{National Institute of Advanced Industrial Science and Technology (AIST), Tsukuba, Ibaraki 305-8568, Japan}

\author{S.~Ishida}
	\affiliation{National Institute of Advanced Industrial Science and Technology (AIST), Tsukuba, Ibaraki 305-8568, Japan}

\author{B.~Thiagarajan}
	\affiliation{Max IV Laboratory, Lund University, Box 118, 22100 Lund, Sweden}

\author{C.~M.~Polley}
	\affiliation{Max IV Laboratory, Lund University, Box 118, 22100 Lund, Sweden}

\author{C.~H.~Lee}
\affiliation{National Institute of Advanced Industrial Science and Technology (AIST), Tsukuba, Ibaraki 305-8568, Japan}

\author{T.~Nishio}
\affiliation{Department of Physics, Tokyo University of Science, Shinjuku, Tokyo 162-8601, Japan}

\author{H.~Eisaki}
	\affiliation{National Institute of Advanced Industrial Science and Technology (AIST), Tsukuba, Ibaraki 305-8568, Japan}
	
\author{I.~Matsuda}
	\affiliation{Institute for Solid State Physics, The University of Tokyo, Kashiwa, Chiba 277-8581, Japan}

%\title{Supplementary Information: \\
	%Enhanced outer-plane superconducting gap in the trilayer cuprate \HgRet}
    
 \title{Supplemental Material: \\
 	Enhanced superconducting gap \red{in the outer CuO$_2$ plane of} the trilayer cuprate \HgRet}

\maketitle

\onecolumngrid

\clearpage

\section{Estimate of Fermi surface area in the small pocket picture}
Fermi surface area of \HgRet\ [(Hg,Re)1223] was estimated based on the doped resonant valence-bond spin-liquid concept~\cite{YangPRB2006}, assuming a small pocket confined between the antiferromagnetic Brillouin-zone (BZ) boundary and the large Fermi surface centered around ($\pi$, $\pi$). While the exact shape of the Fermi surface is difficult to resolve, the area of Fermi surface should fall between the two extreme cases; one is the region enclosed by the large Fermi surface and the antiferromagnetic BZ boundary (indicated in pink in Fig.~S1), and the other is the region enclosed by the large Fermi surface and the mirrored one whose nodal part is tangent to the antiferromagnetic BZ boundary (indicated in green in Fig.~S1). The hole-doping level $p_\mathrm{FS}$ indicated in the main text has been derived by averaging the $p_\mathrm{FS}$ values of the two cases individually for the inner plane (IP) and outer plane (OP).

\begin{figure}[h]
	\begin{center}
		\includegraphics[width=0.7\textwidth]{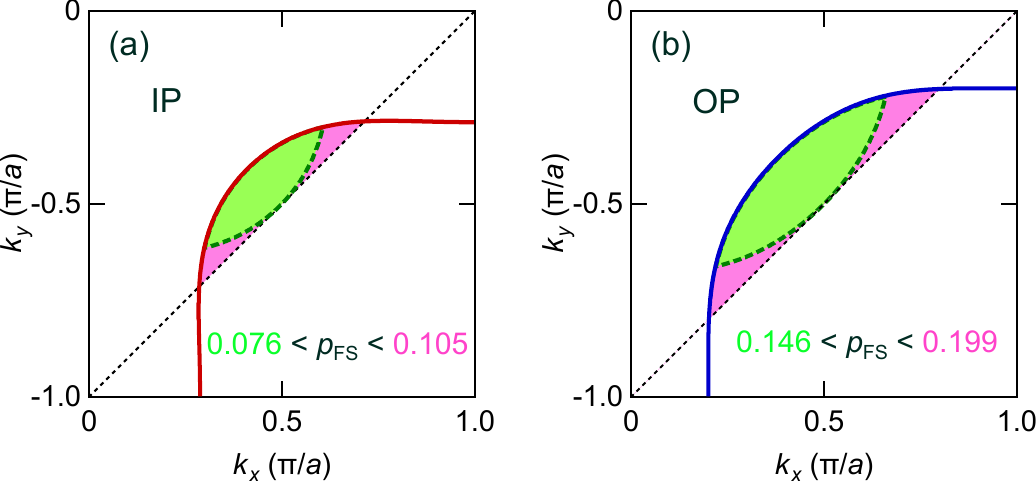}
	\end{center}
	\caption{\textbf{Estimate of Fermi surface area.} Small Fermi surface confined between the large Fermi surface and the antiferromagnetic BZ boundary for the (a) IP and (b) OP. See text for the procedure to estimate the area of the small Fermi surface.} 
	\label{FS_SI}
\end{figure}

\section{Estimate of the energy gap size}
Figure~S2 displays energy distribution curves (EDCs) at the Fermi momentum $k_\mathrm{F}$ before and after symmetrization with respect to the Fermi level \EF\ both for the IP and OP. Quasiparticle peaks are clearly identified and thus the energy gap can be in principle determined by the peak positions of the symmetrized spectra where the influence of the Fermi-Dirac function has been canceled out.  \red{To determine spectral peak positions, the binomial smoothing filter~\cite{Smoothing} was applied. The iteration number of smoothing, $n$, was varied from 20 to 40, and a peak position was extracted from each smoothed spectrum by differentiation. Then, the energy gap $\Delta$ was defined by linearly extrapolating the $n$ dependence of the peak positions to $n=0$, thereby canceling the influence of smoothing. The magnitude of the error bar $\sigma$ can be estimated by $\sigma = \sqrt{\sigma_\mathrm{gap}^2 + \sigma_{h\nu}^2}$, where $\sigma_\mathrm{gap}$ arises from the procedure for extracting the gap size, and $\sigma_{h\nu}$ from the drift of incident photon energy during photoemission measurements. We defined $\sigma_\mathrm{gap}$ as the difference between the estimated $\Delta$ value (spectral-peak energy position linearly extrapolated to $n=0$) and the peak position at $n=40$. $\sigma_{h\nu}$ was fixed at 0.5~meV.}

The evaluation \red{of the gap size} becomes unreliable when the \red{magnitude} is comparable or smaller than the instrumental energy resolution. Such cases are found when the $d$-wave order parameter is smaller than 0.2 in the present case. For those EDCs, which are indicated by red and blue colors in Figs.~S2(a) and (c), respectively, the energy gap $\Delta$ was evaluated from the \red{shift of the} leading-edge \red{midpoint} $\Delta_\mathrm{LEM}$ by employing the empirical relationship of $\Delta \sim 2.2 \Delta_\mathrm{LEM}$~\cite{YoshidaPRL2009}. \red{In this case, $\sigma_\mathrm{gap}$ was determined by inspecting possible leading-edge positions and evaluating the variation of the midpoint. $\sigma_{h\nu}$ was given by 0.5~meV $\times$ 2.2 = 1.1~meV.}
%Note that, even if the gap values estimated from the leading-edge shift are omitted in Fig.~3(n) of the main text, linear extrapolation and hence the estimate of $\Delta_0$ is not significantly affected.

\begin{figure}[h]
	\begin{center}
		\includegraphics[width=\textwidth]{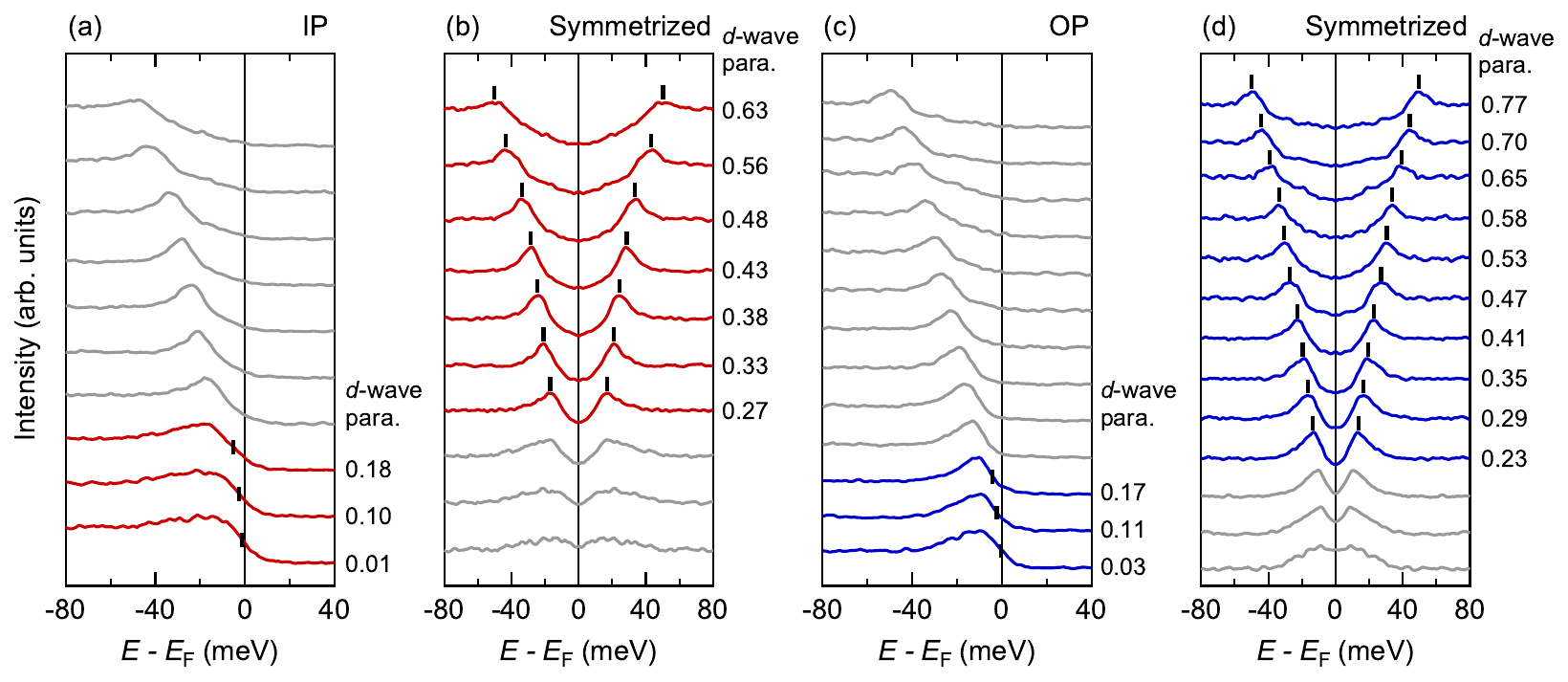}
	\end{center}
	\caption{\textbf{Evaluation of the gap size.} EDCs at $k_\mathrm{F}$ for the IP (a) before and (b) after symmetrization with respect to \EF. The spectra employed to estimate the gap size are highlighted in red. The positions of the leading edge and coherence peak are marked by black bars. (c),(d) Corresponding plots for the OP where blue is used for highlighting.} 
	\label{FS_SI}
\end{figure}

%\bibliography{cuprate}